# Influence of low-level Pr substitution on the superconducting properties of YBa$_2$Cu$_3$O$_{7-\delta}$ single crystals

A Kortyka[1,2], R Puzniak[1], A Wisniewski[1], H W Weber[2], T B Doyle[3], Y Q Cai[4], X Yao[4]

[1] Institute of Physics, Polish Academy of Sciences, PL 02-668 Warsaw, Poland

[2] TU Vienna, Atomic Institute of the Austrian Universities, 1020 Vienna, Austria

[3] School of Physics, University of KwaZulu-Natal, Durban 4000, South Africa

[4] Department of Physics, Shanghai Jiao Tong University, Shanghai 200240, P. R. China

E-mail: bienias@ifpan.edu.pl

**Abstract**. We report on measurements on Y$_{1-x}$Pr$_x$Ba$_2$Cu$_3$O$_{7-\delta}$ single crystals, with $x$ varying from 0 to 2.4%. The upper and the lower critical fields, $H_{c2}$ and $H_{c1}$, the Ginzburg-Landau parameter, $\kappa$, and the critical current density, $J_c(B)$, were determined from magnetization measurements and the effective media approach scaling method. We present the influence of Pr substitution on the pinning force density as well as on the trapped field profiles analyzed by Hall probe scanning.

## 1. Introduction

For practical applications of high-$T_c$ superconductors, it is necessary to improve flux pinning in these materials. Among various methods employed for the enhancement of the pinning force, cationic substitution is known to be effective in increasing the critical current density, $J_c$, significantly [1].

Substitution of three-valence rare earth ions at the Y site in superconducting YBa$_2$Cu$_3$O$_{7-\delta}$ (Y123) does not change the critical temperature, $T_c$, except for Ce, Pr, and Tb. Of particular interest are Pr ions, which destroy superconductivity without changing the structure of Y123 at high substitution levels. They are expected to introduce effective pinning centers with insignificant changes in $T_c$ of the parent compound at low substitution level [2]. Pr-doping on the Y sites in the Y123 structure does not significantly affect the oxygen concentration [3]. Several groups referred to Pr-substituted Y123, but to our knowledge such low-level substitutions have never been seriously considered.

We report on the magnetization and on Hall probe scanning measurements in single-phase monocrystals of the Y$_{1-x}$Pr$_x$Ba$_2$Cu$_3$O$_{7-\delta}$ system. The lower and the upper critical fields, $H_{c1}$ and $H_{c2}$, the thermodynamic critical field, $H_c$, the coherence length, $\xi$, the penetration depth, $\lambda$, and the Ginzburg-Landau parameter, $\kappa = \lambda/\xi$, were estimated. The aim of the studies is to investigate the extent to which low-level Pr substitution influences the superconducting parameters and the flux pinning properties. The critical current density, $J_c(B)$, the pinning force, $F_p$, and the trapped field profiles are presented.

## 2. Experimental details

We investigated $Y_{1-x}Pr_xBa_2Cu_3O_{7-\delta}$ single crystals with small concentration of Pr: $x = 0$, 0.013, and 0.024 grown by top seeded solution growth [4]. The crystals were plate-shaped with dimensions 4.3×4.1×1 mm³, 5.2×3.7×1.2 mm³, and 4.7×3.35×2.1 mm³ for $x$ = 0, 0.013, and 0.024, respectively. The crystals were annealed in flowing oxygen at 500 ºC for 72 hours followed by furnace cooling to room temperature. The Pr content was determined by an inductively coupled plasma (ICP) technique.

The dc magnetization measurements were made in a 7 T SQUID magnetometer (Quantum Design, MPMS) and a 9 T Physical Property Measurement System (Quantum Design, PPMS) with the magnetic field $H \parallel c$. The scan length of the specimen in the SQUID magnetometer was 3 cm. The field dependence of the magnetization was measured. By extrapolating the reversible part in the logarithmic $M(\ln H)$ dependence to zero, the values of $H_{c2}(T)$ were estimated [5]. The normal state signal was subtracted by fitting Curie law parameters to the temperature dependence of the magnetization recorded at temperatures between 130 and 300 K. The $J_c$ values were calculated using the modified Bean model [6].

The superconducting state parameters were determined by fitting free parameters in the theoretical magnetization dependence based on Ginzburg-Landau (GL) theory, in order to obtain the reversible magnetization. The effective media approach to magnetization scaling [7] was proved to be successful for high and low temperature superconductors and was applied here.

In order to analyze the trapped field profile, Hall probe scanning was used. The sample was magnetized in a field of 0.4 T. The distance between the sample and the Hall probe was 0.1 mm. The measurements were performed in a liquid nitrogen bath.

## 3. Results and discussion

Zero- and field-cooled magnetization as a function of temperature in a field of 0.5 mT is presented in Figure 1 for various Pr concentrations. All single crystals have superconducting-to-normal state transition widths of around 1 K and the same $T_c = 90.1 \pm 0.2$ K. With increasing Pr content, a continuous progression of the fishtail effect is observed (see insert of Figure 1).

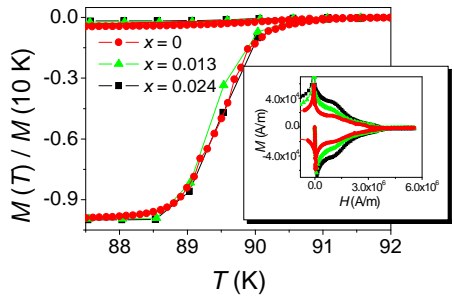
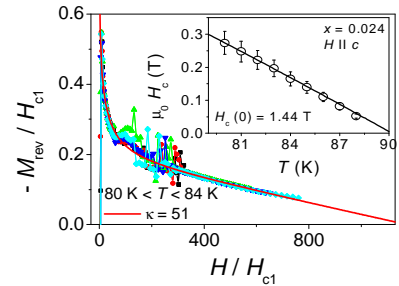

Figure 1. Normalized temperature dependence of the magnetization for $Y_{1-x}Pr_xBa_2Cu_3O_{7-\delta}$ single crystals. Insert: field dependence of the magnetization for all crystals at $T = 80$ K.

Figure 2. Magnetization scaling for $Y_{0.976}Pr_{0.024}Ba_2Cu_3O_{7-\delta}$. The scatter of the experimental points below the reversible region is expected and related to the fishtail effect. Insert: fit of the theoretical expression (given in the text) to the thermodynamic critical field values.

Scaling of the experimental reversible magnetization according to the theory is presented in Figure 2. The scaling method requires only two parameters: the sample dimensions and the intrinsic GL parameter $\kappa$. By fitting the initial slope of the theoretical reversible magnetization to that obtained experimentally, $H_{c1}$ was extracted. Then $H_{c2}$, $H_c$, $\xi$, and $\lambda$ were estimated using simple relations between the superconducting parameters.

The results for $H_{c2}(T)$ derived from the extrapolation of the reversible magnetization are consistent with those obtained from the magnetization scaling method. $dH_{c2}/dT$ is -1.83, -2.00, and -2.09 T/K for $x$ = 0, 0.013, and 0.024, respectively. Using the Werthamer-Helfand-Hohenberg (WHH) approximation [8], $\mu_0 H_{c2}(0)$ is estimated to be 119.2, 130.5, and 136.8 T for $x = 0$, 0.013, and 0.024, respectively. Low-level Pr substitution is not expected to significantly change the intrinsic properties of a superconductor, which is consistent with a constant value of the GL parameter, within experimental error, obtained here. The values of $\mu_0 H_c(0)$ (see insert of Figure 2) were derived from the relation $\dfrac{H_c(T)}{H_c(0)} = 1.7367\left[1-\dfrac{T}{T_c}\right]\left[1-0.2730\left(1-\dfrac{T}{T_c}\right)-0.0949\left(1-\dfrac{T}{T_c}\right)^2\right]$, [9], and were found to be 1.33, 1.38, and 1.44 T for $x = 0$, 0.013, and 0.024, respectively.

A summary of the results for the thermodynamic parameters is given in Table 1.

Table 1. Thermodynamic parameters of $Y_{1-x}Pr_xBa_2Cu_3O_{7-\delta}$ monocrystals.

| Pr content | $T_c$ (K) | $\mu_0 H_{c1}^c(0)$ (mT) | $\mu_0 H_{c2}^c(0)$ (T) | $\mu_0 H_c^c(0)$ (T) | $\xi_{ab}(0)$ (nm) | $\lambda_{ab}(0)$ (nm) | $\kappa_c$ |
|---|---|---|---|---|---|---|---|
| 0 | 90.1 | 96.6 | 119.2 | 1.33 | 1.66 | 81.5 | 49 |
| 0.013 | 89.9 | 102 | 130.5 | 1.38 | 1.59 | 79.4 | 50 |
| 0.024 | 90.2 | 103 | 136.8 | 1.44 | 1.55 | 79.1 | 51 |

The critical current densities derived from the modified Bean model are presented in Figure 3. Although $H_{irr}(T)$ does not change at low-level substitution, $J_c(B)$ increases in the substituted samples. The lack of a systematic increase is probably due to disorder from oxygen deficient regions, which is not well known. As pointed out by many authors, the critical current density strongly depends on that.

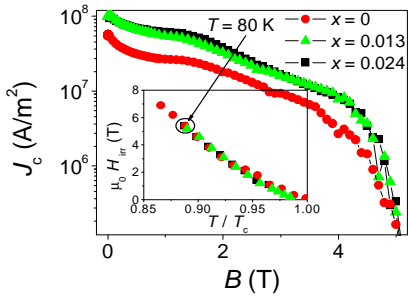
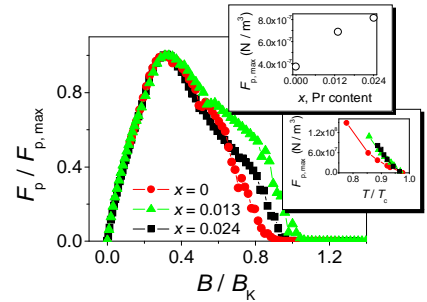

Figure 3. Increase of the critical current density with Pr substitution. Insert: irreversibility lines.

Figure 4. Normalized pinning force density for $Y_{1-x}Pr_xBa_2Cu_3O_{7-\delta}$ at $T = 80$ K. Upper insert: increase of $F_{p,\,max}$ with increasing Pr substitution. Lower insert: temperature dependence of $F_{p,\,max}$.

Valuable information on the pinning mechanisms can be obtained from the functional form of the pinning force density, $F_p(B) = J_c B$. All the temperature-dependent curves of $F_p/F_{p,\,max}$ versus $b$ ($b = B/B_k$, where $B_k$ is defined from Kramer's plots and is related to the irreversibility field rather than to the upper critical field) would collapse to a single curve, if the primary pinning mechanism was not changed. In Y123 crystals, oxygen-deficient regions act as pinning centers. In the Pr doped samples, besides oxygen vacancies, defects introduced by the magnetic ions are present. Additional defects, like small cracks or twin boundaries, naturally introduced during the oxygenation process, are expected. With increasing Pr content, no pronounced shift of the field, where $F_{p,\,max}$ occurs, was observed (Figure 4). This, together with the presence of a single curve at low fields, suggests that the pinning centers are of the same nature and equally strong up to 2.4% of Pr substitution. Therefore, pinning is assumed to be mainly due to oxygen deficient regions at low fields. At high fields, on the other hand, the pinning force density strongly depends on the Pr content. We expect, that close to the irreversibility line only the strongest pinning centers remain active. They are probably magnetic ions together with large oxygen deficient regions. Since the oxygen content may differ in the crystals, a difference in the normalized pinning force density is observed. A similar behavior was observed over a wide range of temperatures. The emergence of a peak in the magnetization with increasing field at high temperatures suggests that the pinning centers are field-induced.

The influence of the Pr substitution on the trapped field profiles was analyzed by Hall probe scanning and is presented in Figure 5. An increase in the trapped field, $B_t$, is observed in the Pr-substituted crystals. These results are consistent with the $J_c(B)$ increase.

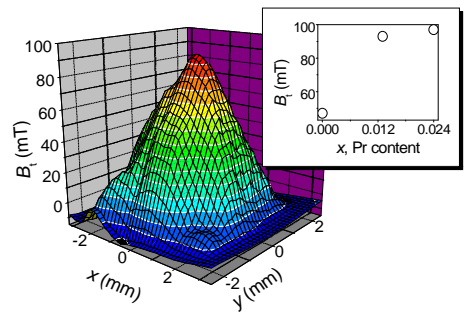

Figure 5. Hall probe scanning profile of $Y_{0.987}Pr_{0.013}Ba_2Cu_3O_{7-\delta}$ single crystal. Insert: increase of the trapped field in Pr-substituted crystals.

## 4. Conclusions

We report on the effect of low-level Pr substitution on the superconducting parameters for $H \parallel c$-axis of large-sized single crystals of $Y_{1-x}Pr_xBa_2Cu_3O_{7-\delta}$. For crystals with $x \leq 2.4\%$ and unchanged $T_c$, we find an increase of $H_{c2}$ as well as of $H_c$ with increasing Pr substitution. No change in the GL parameter, $\kappa$, is observed and its value found to be equal to $\kappa = 50 \pm 3$. By increasing the Pr content, higher $J_c(T, H)$ values are obtained, while no apparent dependence in $H_{irr}(T)$ was observed. The results indicate that the substitution of Pr in Y123 induces effective pinning centers, enhances the pinning force and the trapped field by a factor of about 2 at 77 K.

Acknowledgments
A.K. thanks the European NESPA project for financial support. This work was partially supported by the Polish Ministry of Science and Higher Education under grant No. N N202 4132 33.